# Self - Organized Si Dots On Ge Substrates

D. Pachinger, H. Lichtenberger und F. Schäffler

*Institute of Semiconductor and Solid State Physics, Johannes Kepler University Linz,
Altenbergerstr. 69, A-4040 Linz, Austria*

**Abstract** The epitaxial growth conditions for silicon on germanium substrates were investigated as a function of growth temperature and monolayer coverage. Island formation was observed for the hole studied temperature range, although strong alloying with the substrate occurred for the highest temperatures. Carbon pre-deposition offers suitable nucleation centers for the Si island and reduction of alloying. pre-structured Ge substrates were prepared to enhance islanding and to achieve ordering.

## INTRODUCTION

In the Si/SiGe heterosystem self-organization schemes based on Stranski–Krastanov (SK) growth were mainly investigated for compressively strained Ge layers on Si substrates, which offer hole confinement in the islands. Due to the Type II band alignment, electron confinement requires tensely strained, Si-rich dots, which can be realized on Ge substrates or pseudosubstrates. Very little is known about SK growth of Si on Ge, but there are strong indications that dot formation is kinetically hampered, if the epilayer is under tensile strain [1]. On the other hand, SK growth is driven by total energy minimization, and since the elastic energy depends quadratically on strain, close to thermal equilibrium 3D island formation is not expected to depend on the sign of the lattice mismatch. Indeed, SK growth has already been observed in layers of IV-VI heterostructures under tensile strain [2].

Here, we report on MBE growth conditions for Si island formation on Ge(001) substrates. Buffer layer growth was optimized and Si island formation was investigated in the temperature range between 550 and 750 °C for 5 to 15 ML Si coverage. Seeded nucleation was observed after carbon pre-deposition on a Ge buffer layer. Finally, Ge substrates were pre-structured to offer suitable nucleation sites for the Si islands.

## EXPERIMENTAL

Substrates used in this study were Ga-doped Cz-Ge(100) wafers with a resistivity of about 8 Ωcm. Pieces of 17,5x17,5 mm were chemically pre-cleaned [3] and loaded into our Riber SIVA45 MBE system via a load lock system, followed by an in-situ thermal oxide desorption step at 750 °C for 15 minutes.

AFM measurements were carried out for the characterization of buffer- and island growth. Surface orientation maps extracted from these AFM images were used for facet analysis.

Pre-structuring of the Ge substrates was realized by e-beam lithography with a Leo Supra 35 FE-SEM at 20 kV and subsequent reactive ion etching in an Oxford Plasmalab 80 reactor with 100 % $CF_4$. (50 sccm $CF_4$ flow, 30 W, 50 mTorr pressure) The etch rate was about 100 nm/min for the chosen parameters.

## RESULTS AND DISCUSSION

After oxide desorption a Ge buffer layer was deposited. By systematically varying its deposition temperature, we found that a 50 nm thick Ge buffer grown at 400 °C offers the lowest mean-root-square roughness of about 0.1 nm (figure 1).

On this buffer we deposited 5 - 15 ML of Si at temperatures between 550 °C and 750 °C. Island formation was observed at the highest deposition temperature (figure 2A), but these islands were almost complete alloyed with the Ge substrate. Lowering the growth temperature to 550 °C leads to a drastic reduction of the dot density and to a pronounced trench around the islands (figure 2B).

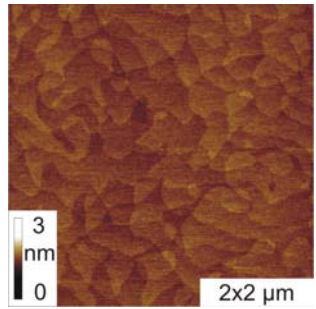

**FIGURE 1.** AFM image of a 50 nm thick Ge buffer layer on a Ge (001) substrate grown at 400 °C providing a mean-root-square roughness of about 0.1 nm. Flat areas seen in the image are separated by two-atomic height steps.

These experiments confirm the kinetically restricted nucleation of Si islands at lower growth temperatures, which is, in our case, the only temperature range where alloying with the buffer layer can be efficiently suppressed. It is therefore essential to enhance island formation in tensile strained films by providing suitable nucleation centers. For this purpose we deposited a fraction of a ML of carbon onto the Ge buffer layer just prior to Si deposition. With carbon pre-deposition varying from 0.05 to 0.5 ML, Si island growth could be observed over the whole range of growth temperatures investigated (figure 3).

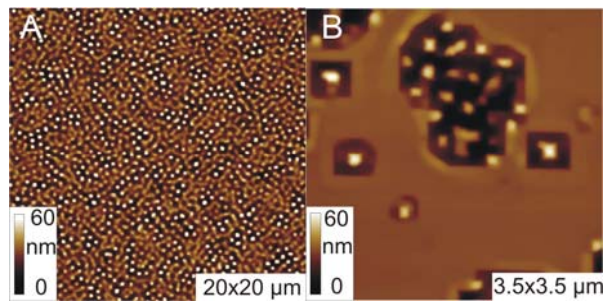

**FIGURE 2.** A: AFM images of 10 ML Si grown on a 50 nm Ge buffer layer at 750 °C showing a dot density of $4 \times 10^8$ 1/cm²; B: AFM images of 10 ML Si grown on a 50 nm Ge buffer layer at 675 °C, showing coupled and uncoupled trenches enclosing the dots; dot density is $3 \times 10^8$ 1/cm²

Extracting surface orientation maps from the AFM images we found for both growth conditions Si-rich islands with {105}, {113} and {15 3 23} facets (figure 3B). These are the same characteristic facets found for SiGe islands on Si, indicating that Si also forms the well-known dome-shaped and pyramid-shaped islands [4]. Especially the appearance of the {105} facet is somewhat surprising, because it is usually associated with a facet that appears only on compressively strained SiGe and Ge layers.

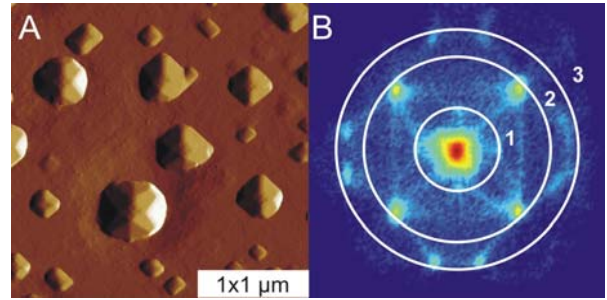

**FIGURE 3.** A: AFM image in the derivative mode of 10 ML Si grown on Ge (001) substrate at 675 °C; B: Surface orientation map of A, showing 1:{105}, 2:{113} and 3:{15 3 23} facets, which are the same characteristic facets found for dome- and pyramid-shaped Ge dots on Si.

As a further step to enhance island nucleation and order, pit- and trench-structures with a periodicity of 800 and 400 nm were realized by reactive ion etching with $CF_4$. The overgrowth of the pit-structures with a 500 nm Ge buffer layer at a temperature of 400 °C leads to pits of about 80 nm depth showing {105}, {113} and {15 3 23} facets (figure 4A and 4B).

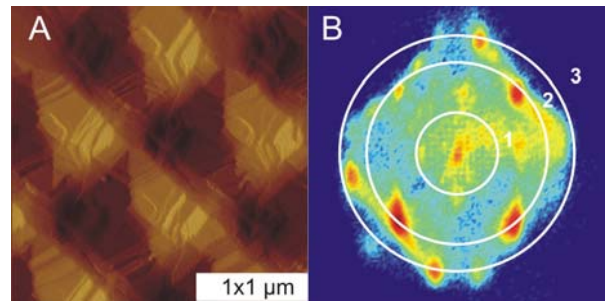

**FIGURE 4.** A: AFM image in the derivative mode of a 50 nm thick Germanium buffer layer on a pre-structured Ge (001) substrate grown at 400 °C; B: Surface Orientation Map of A, showing 1:{105}, 2:{113} and 3:{15 3 23} facets.

On these templates we already observed preferential Si island formation, indicating that these structures offer suitable nucleation centers.

## ACKNOWLEDGMENTS

Financial support from the FWF (Vienna) via SFB-IRON, and from GMe are gratefully acknowledged.